\documentclass{crmp-l}
\copyrightinfo{2003}{Willy Hereman}

\newtheorem{theorem}{Theorem}[section]

\theoremstyle{definition}
\newtheorem{definition}[theorem]{Definition}
\newtheorem{example}[theorem]{Example}

\theoremstyle{remark}
\newtheorem{remark}[theorem]{Remark}

\numberwithin{equation}{section}

\DeclareMathOperator{\rank}{rank}

\begin{document}

\title[Densities, symmetries, and recursion operators for nonlinear DDEs]
{Symbolic Computation of Conserved Densities, Generalized Symmetries, 
and Recursion Operators for Nonlinear Differential-Difference Equations} 


\author{Willy Hereman}
\address{Department of Mathematical and Computer Sciences, 
Colorado School of Mines, Golden, Colorado 80401-1887}
\email{whereman@mines.edu}
\thanks{This material is based upon work supported by the National Science 
Foundation under Grant No.\ CCR-9901929.}
%
%
%
\author{Jan A.\ Sanders}
\address{
Department of Mathematics, 
Faculty of Sciences, 
Free University, De Boelelaan 1081a,
1081 HV Amsterdam, The Netherlands}
\email{jansa@cs.vu.nl}
%

\author{Jack Sayers}
\address{
Physics Department 103-33,
California Institute of Technology,
Pasadena, California 91125}
\email{jack@its.caltech.edu}
%

\author{Jing Ping Wang}
\address{
Department of Mathematics,
Brock University, 
St.\ Catharines, Ontario, Canada L2S 3A1}
\email{jwang@spartan.ac.BrockU.ca}
%

\date{August 31, 2003.}
\dedicatory{This paper is dedicated to Ryan Sayers (1982-2003)}

\begin{abstract}
Algorithms for the symbolic computation of conserved densities, fluxes, 
generalized symmetries, and recursion operators for systems of 
nonlinear differential-difference equations are presented. 
In the algorithms we use discrete versions of the Fr\'echet and 
variational derivatives, as well as discrete Euler and homotopy operators.

The algorithms are illustrated for prototypical nonlinear lattices, 
including the Kac-van Moerbeke (Volterra) and Toda lattices.
Results are shown for the modified Volterra and Ablowitz-Ladik lattices.
\end{abstract}
\maketitle
\section{Introduction}
\label{introduction}
The study of complete integrability of nonlinear differential-difference 
equations (DDEs) largely parallels that of PDEs 
\cite{MAandPC91,VAandASandRY00,JW98}. 
Indeed, as in the continuous case, the existence of sufficiently many
conserved densities and generalized symmetries is a predictor for complete 
integrability.
Based on the first few densities and symmetries a recursion operator 
(which maps symmetries to symmetries) can be constructed.  
The existence of a recursion operator, which allows one to generate an 
infinite sequence of symmetries, confirms complete integrability.

There is a vast body of work on the complete integrability of DDEs. 
Consult e.g.\ \cite{VAandASandRY00,MHandWH03} for various approaches and
references.
In this article we describe algorithms to symbolically compute polynomial 
conservation laws, fluxes, generalized symmetries, and recursion operators 
for DDEs.
The design of these algorithms heavily relies on related work for PDEs
\cite{UGandWH97a,WHandUG99,WHandUGandMCandAM98} and work by Oevel 
{\it et al} \cite{WOandHZandBF89}.

There exists a close analogy between the continuous and discrete (in space)
cases. 
As shown in \cite{PHandLM02,LMandPH02}, this analogy can be completely 
formalized and both theories can be formulated in terms of complexes.
The same applies for the formulation in terms of Lie algebra 
complexes in \cite{JW98}.
This allows one to translate by analogy the existing algorithms immediately
(although complications arise when there is explicit dependence
on the space variable in the discrete case).
One of the more useful tools following from the abstract theory
is the homotopy operator, which is based on scaling vectorfields,
and goes back to Poincar{\'e} in the continuous case. 
This operator allows one to directly integrate differential forms
and can be straightforwardly implemented in computer algebra packages,
since it reduces to integration over one scaling parameter.
The discrete analogue as described in \cite{PHandLM02,LMandPH02} does the 
corresponding job in the discrete case and we use it to compute fluxes.
In this paper we do not explicitly use the abstract framework,
yet, it has been a motivating force for the development of our algorithms.


The algorithms in this paper can be implemented in any computer 
algebra system. 
Our Mathematica package {\it InvariantsSymmetries.m} \cite{UGandWH97code} 
computes densities and symmetries, and therefore aids 
in automated testing of complete integrability of semi-discrete lattices.
Mathematica code to automatically compute recursion operators is still
under development.

The paper is organized as follows. 
In Section~\ref{definitions}, we give key definitions and introduce 
the Kac-van Moerbeke (KvM) \cite{MKandPvM75}, Toda \cite{MT81} and 
Ablowitz-Ladik (AL) lattices \cite{MAandJL75}, 
which will be used as prototypical examples throughout the paper. 
The discrete higher Euler operators (variational derivatives) and the 
discrete homotopy operator are introduced in Section~\ref{discreteeuler}.
These operators are applied in the construction of densities and
fluxes in Section~\ref{algodensityflux}.
The algorithm for higher-order symmetries is outlined in 
Section~\ref{algosymmetry}. 
The algorithms for scalar and matrix recursion operators are given in 
Section~\ref{algorecursionscalar} and Section~\ref{algorecursionmatrix}.
The paper concludes with two more examples in Section~\ref{examples}.
%
%
%
%
%
\section{Key Definitions and Prototypical Examples}
\label{definitions}

\begin{definition}
A nonlinear (autonomous) DDE is an equation of the form 
\begin{equation}
\label{dde}
{\dot{\bf u}}_n = 
{\bf F} (...,{\bf u}_{n-1}, {\bf u}_{n}, {\bf u}_{n+1},...),
\end{equation}
where ${\bf u}_{n}$ and ${\bf F}$ are vector-valued functions with $m$ 
components.
The integer $n$ corresponds to discretization in space; the dot denotes 
one derivative with respect to the continuous time variable $(t).$
\end{definition}
For simplicity, we will denote the components of ${\bf u}_n$ by 
$(u_n, v_n, w_n, \cdots).$ 
For brevity, we write ${\bf F}({\bf u}_n),$ although ${\bf F}$ 
typically depends on ${\bf u}_n$ and a finite number of its forward and 
backward shifts.
We assume that ${\bf F}$ is polynomial with constant coefficients.
If present, parameters are denoted by lower-case Greek letters.
No restrictions are imposed on the level of the forward or backward 
shifts or the degree of nonlinearity in ${\bf F}.$
\begin{example}
The AL lattice \cite{MAandJL75}, 
\begin{eqnarray}
\label{ablowitzladiklattice}
{\dot{u}}_n 
&=& ( u_{n+1} - 2 u_n + u_{n-1} ) + u_n v_n (u_{n+1} + u_{n-1}), 
\nonumber \\
{\dot{v}}_n 
&=& - ( v_{n+1} - 2 v_n + v_{n-1} ) -  u_n v_n (v_{n+1} + v_{n-1}), 
\end{eqnarray}
is a completely integrable discretization of the nonlinear Schr\"odinger 
equation.
\end{example}
%
%
\begin{definition}
A DDE is said to be dilation invariant if it is invariant under a scaling 
(dilation) symmetry.  
\end{definition}
\begin{example}
The KvM lattice \cite{MKandPvM75},
\begin{equation}
\label{kvmlattice}
{\dot {u}}_n = u_n (u_{n+1} - u_{n-1}),
\end{equation}
is invariant under 
$(t, u_n) \rightarrow (\lambda^{-1} t, \lambda u_n), $
where $\lambda$ is an arbitrary scaling parameter.
\end{example}
\begin{example}
The Toda lattice \cite{MT81} in polynomial form \cite{UGandWH98},
\begin{equation}
\label{todalattice}
{\dot{u}}_n = v_{n-1} - v_n, \quad {\dot{v}}_n = v_n (u_n - u_{n+1}),
\end{equation}
is invariant under the scaling symmetry
\begin{equation}
\label{todascale}
(t, u_n, v_n) \rightarrow (\lambda^{-1} t, \lambda u_n, {\lambda}^2 v_n).
\end{equation}
Thus, $u_n$ and $v_n$ correspond to one, respectively two derivatives with 
respect to $t,$
\begin{equation}
\label{todascaleweights}
u_n \sim \frac{\rm{d}}{\rm{d}t}, \quad 
v_n \sim \frac{{\rm{d}}^2}{{\rm{d}t}^2}.
\end{equation}
\end{example}
\begin{definition}
We define the weight, $w$, of a variable as the number of $t\/$-derivatives 
the variable corresponds to.
\end{definition}
Since $t$ is replaced by $\frac{t}{\lambda},$ we set 
$w(\frac{\rm{d}}{\rm{d}t})=1.$ 
In view of (\ref{todascaleweights}), we have $w(u_n)=1$, and $w(v_n)=2$
for the Toda lattice.
Weights of dependent variables are nonnegative, rational, and 
independent of $n$.
\begin{definition}
The rank of a monomial is defined as the total weight of the monomial.
An expression is uniform in rank if all its terms have the same rank.
\end{definition}
\begin{example}
In the first equation of (\ref{todalattice}), all the monomials have
rank 2; in the second equation all the monomials have rank 3. 
Conversely, requiring uniformity in rank for each equation in 
(\ref{todalattice}) allows one to compute the weights of the dependent 
variables with simple linear algebra. 
Indeed,
\begin{equation}
\label{todaweightequations}
w(u_n) + 1 \!=\! w(v_n), \quad w(v_n) + 1 \!=\! w(u_n) + w(v_n), 
\end{equation}
yields
\begin{equation}
\label{todaweights}
w(u_n) = 1, \quad w(v_n) = 2, 
\end{equation} 
which is consistent with (\ref{todascale}).
\end{example}
Dilation symmetries, which are special Lie-point symmetries, are common 
to many lattice equations. 
Lattices that do not admit a dilation symmetry can be made scaling invariant
by extending the set of dependent variables using auxiliary parameters 
with scaling. 
\begin{example}
The AL lattice is not dilation invariant.
Introducing an auxiliary parameter $\alpha,$ hence replacing
(\ref{ablowitzladiklattice}) by 
\begin{eqnarray}
\label{ablowitzladiklatticenew}
{\dot{u}}_n 
&=& \alpha ( u_{n+1} - 2 u_n + u_{n-1} ) + u_n v_n (u_{n+1} + u_{n-1}), 
\nonumber \\
{\dot{v}}_n 
&=& - \alpha ( v_{n+1} - 2 v_n + v_{n-1} ) -  u_n v_n (v_{n+1} + v_{n-1}), 
\end{eqnarray}
and requiring uniformity in rank, gives
\begin{eqnarray}
\label{ablowitzladikweightequations}
w(u_n) + 1 &\!=\!& w(\alpha) + w(u_n) = 2 w(u_n) + w(v_n), 
\nonumber \\
w(v_n) + 1 &\!=\!& w(\alpha) + w(v_n) = 2 w(v_n) + w(u_n).
\end{eqnarray}
Obviously,
\begin{equation}
\label{ablowitzladikweighteq1}
w(u_n) + w(v_n) = w(\alpha) = 1.
\end{equation} 
Several scales are possible. 
The choice $w(u_n) = w(v_n) = \frac{1}{2}, w(\alpha) = 1,$ corresponds 
to the scaling symmetry
\begin{equation}
\label{ablowitzladikscale}
(t, u_n, v_n, \alpha) 
\rightarrow (\lambda^{-1} t, \lambda^{\frac{1}{2}} u_n, 
{\lambda}^{\frac{1}{2}} v_n, \lambda \alpha ).
\end{equation}
\end{example}
%
%
%
\begin{definition}
A scalar function $\rho_n({\bf u}_n)$ is a conserved density of (\ref{dde}) 
if there exists a scalar function $J_n({\bf u}_n)$ called the associated flux, 
such that 
\begin{equation} 
\label{ddeconslaw} 
{\rm D}_t \rho_n + \Delta \, J_n = 0 
\end{equation}
is satisfied on the solutions of (\ref{dde}) \cite{PO93}. 
\end{definition}
In (\ref{ddeconslaw}), we used the (forward) difference operator, 
$\Delta = {\rm D} - {\rm I},$ 
defined by 
\begin{equation}
\label{delta}
\Delta \, J_n = ({\rm D} - {\rm I}) \, J_n = J_{n+1} - J_n,
\end{equation}
where ${\rm D}$ denotes the up-shift (forward or right-shift) operator, 
${\rm D} J_n = J_{n+1},$ and $\rm I$ is the identity operator. 
Operator $\Delta$ takes the role of a spatial derivative on the shifted 
variables as many DDEs arise from discretization of a PDE in 
$(1+1)$ variables. 
Most, but not all, densities are polynomial in ${\bf u}_n.$
%
%
%
\begin{example}
The first three conservation laws for (\ref{kvmlattice}) are
\begin{eqnarray}
\label{kvmconslaw1}
& & {\rm D}_{t} (\ln(u_{n}))
+ (u_{n+1} + u_{n}) - (u_{n} + u_{n-1}) = 0, \\
\label{kvmconslaw2}
& & {\rm D}_{t}(u_n)
+ (u_{n+1} u_n) - (u_n u_{n-1}) =0,  \\
\label{kvmconslaw3}
& &  {\rm D}_{t}(\frac{1}{2} u_n^2 + u_n u_{n+1})
+ u_{n}u_{n+1}(u_{n+1} + u_{n+2}) - u_{n-1}u_{n}(u_n + u_{n+1}) = 0.
\end{eqnarray}
\end{example}
%
%
\begin{example}
For the Toda lattice (\ref{todalattice}) the first four density-flux pairs are
\begin{eqnarray}
\label{condenstoda0}
\rho_n^{(0)} \!&\!=\!&\! \ln (v_n), 
\;\;\;\quad\quad\quad\quad \quad\quad\quad\quad
J_n^{(0)} = u_{n}, \\
\label{condenstoda1}
\rho_n^{(1)} \!&\!=\!&\! u_n, 
\;\;\;\;\;\quad \quad \quad \quad \quad  \quad\quad\quad\quad
J_n^{(1)} = v_{n-1}, \\
\label{condenstoda2}
\rho_n^{(2)} \!&\!=\!&\! \frac{1}{2} u_n^2 + v_n, 
\;\;\;\;\;\quad \quad  \quad\quad\quad\quad
J_n^{(2)} = u_n v_{n-1}, \\
\label{condenstoda3}
\rho_n^{(3)}  \!&\!=\!&\! \frac{1}{3} {u_n}^3 + u_n ( v_{n-1} + v_n ),
\;\;\;\quad J_n^{(3)} = u_{n-1} u_n v_{n-1} + {v_{n-1}}^2.
\end{eqnarray}
\end{example}
The above densities are uniform of ranks $0$ through $3.$
The corresponding fluxes are also uniform in rank with ranks $1$ through $4.$
In general, if in (\ref{ddeconslaw}) ${\rm rank}\, \rho_n=R$ 
then ${\rm rank}\, J_n=R+1,$ since $w({\rm D}_t) = 1.$ 

This comes as no surprise since the conservation law (\ref{ddeconslaw})
holds on solutions of (\ref{dde}), hence it `inherits' the dilation 
symmetry of (\ref{dde}).
%

In Section~\ref{algodensityflux} we will give an algorithm to compute
polynomial conserved densities and fluxes and use (\ref{todalattice}) 
to illustrate the steps.
Non-polynomial densities (which are easy to find by hand) can be computed 
with the method given in \cite{MHandWH03}.
\begin{definition}
Compositions of ${\rm D}$ and ${\rm D}^{-1}$ define an equivalence relation 
$(\equiv)$ on monomials in the components of ${\bf u}_n.$
All shifted monomials are equivalent.
\end{definition}
\begin{example}
For example, 
$u_{n-1} v_{n+1} \, \equiv \, u_{n+2} v_{n+4} \, \equiv \, u_{n-3} v_{n-1}.$
\end{example}
%
%
\begin{definition}
The main representative of an equivalence class is the monomial of the 
class with $n$ as lowest label on any component of $u$ 
(or $v$ if $u$ is missing).
\end{definition}
\begin{example} 
For example, $u_n u_{n+2}$ is the main representative of the class 
$\{\cdots,u_{n-2} u_{n},u_{n-1} u_{n+1},u_n u_{n+2},u_{n+1} u_{n+3},\cdots\}.$ 
Use lexicographical ordering to resolve conflicts. 
For example, $u_n v_{n+2} $ (not $u_{n-2} v_n$) is the main representative 
of the class $\{\cdots,u_{n-3} v_{n-1},\cdots,u_{n+2} v_{n+4},\cdots \}.$ 
\end{example}
%
%
\begin{definition}
A vector function ${\bf G}({\bf u}_n)$ is called a generalized symmetry 
of (\ref{dde}) if the infinitesimal transformation 
${\bf u}_n \rightarrow {\bf u}_n + \epsilon {\bf G}$ 
leaves (\ref{dde}) invariant up to order $\epsilon.$ 
Consequently, ${\bf G}$ must satisfy \cite{PO93}
\begin{equation}
\label{ddesymmetry}
{\rm D}_{t}{\bf G} = {\bf F}'({\bf u}_n)[{\bf G}],
\end{equation}
on solutions of (\ref{dde}). 
$\!{\bf F}'({\bf u}_n)[{\bf G}]$ is the Fr\'echet derivative of 
${\bf F}$ in the direction of $\!{\bf G}.$ 
\end{definition}
For the scalar case $(N=1)$, the Fr\'echet derivative in the direction of 
$G$ is 
\begin{equation}
\label{ddefrechetscalar}
F'(u_n)[G] = \frac{\partial}{\partial{\epsilon}}
F(u_n +\epsilon G) {|_{\epsilon = 0}}
= \sum_{k} \frac{\partial F}{\partial u_{n+k}} {\rm D}^{k} G,
\end{equation}
which defines the Fr\'echet derivative operator
\begin{equation}
\label{ddefrechetscalaroperator}
F'(u_n) = \sum_{k} \frac{\partial F}{\partial u_{n+k}} {\rm D}^{k}.
\end{equation}
In the vector case with two components, $u_n$ and $v_n,$ the Fr\'echet 
derivative operator is 
\begin{equation}
\label{ddefrechetvectoroperator}
{\bf F}'({\bf u}_n)  
= \left( \begin{array}{cc} 
\sum_{k} \frac{\partial F_1}{\partial u_{n+k}} {\rm D}^{k} & 
\sum_{k} \frac{\partial F_1}{\partial v_{n+k}} {\rm D}^{k} \\ 
& \\
\sum_{k} \frac{\partial F_2}{\partial u_{n+k}} {\rm D}^{k} & 
\sum_{k} \frac{\partial F_2}{\partial v_{n+k}} {\rm D}^{k}
\end{array} \right).
\end{equation}
Applied to ${\bf G} = (G_1 \quad G_2)^{\rm T},$
where $T$ is transpose, one obtains
\begin{equation}
\label{ddefrechetcomponent}
{F_i}'({\bf u}_n)[{\bf G}] = 
\sum_{k} \frac{\partial F_i}{\partial u_{n+k}} {\rm D}^{k} G_1
+ \sum_{k} \frac{\partial F_i}{\partial v_{n+k}} {\rm D}^{k} G_2, \quad i=1,2.
\end{equation}
In (\ref{ddefrechetscalar}) and (\ref{ddefrechetcomponent}) 
summation is over all positive and negative shifts (including $k=0).$
For $k>0$, 
${\rm D}^{k} = {\rm D} \circ {\rm D} \circ \cdots \circ {\rm D} \; (k$ times). 
Similarly, for $k<0$ the down-shift operator ${\rm D}^{-1}$ is applied 
repeatedly.
The generalization of (\ref{ddefrechetvectoroperator}) to a system with 
$N$ components should be obvious.
%
%
\begin{example}
The first two symmetries of (\ref{kvmlattice}) are
\begin{eqnarray}
\label{kvmsymm1}
G^{(1)} &=& u_n (u_{n+1}-u_{n-1}), \\ 
\label{kvmsymm2}
G^{(2)} &=& 
u_n u_{n+1}  (u_n + u_{n+1} + u_{n+2}) 
- u_{n-1} u_n (u_{n-2} + u_{n-1} + u_n).
\end{eqnarray}
\end{example}
These symmetries are uniform in rank (rank 2 and 3, respectively).
The symmetries of ranks 0 and 1 are both zero.
\begin{example}
The first two non-trivial symmetries of (\ref{todalattice}) are
\begin{equation}
\label{todasymm1}
{\bf G}^{(1)} 
= \left( \begin{array}{c} 
v_{n-1}-v_{n} \\ 
v_{n}(u_{n}-u_{n+1}) 
\end{array} \right), 
\end{equation}
and
\begin{equation}
\label{todasymm2}
{\bf G}^{(2)} 
= \left( \begin{array}{c} 
v_{n}(u_{n} + u_{n+1}) - v_{n-1}(u_{n-1}+u_{n}) \\
v_{n}(u_{n+1}^{2} - u_{n}^{2} + v_{n+1} - v_{n-1})
\end{array} \right).
\end{equation}
\end{example}
%
The above symmetries are uniform in rank. 
For example, ${\rank}\,G_1^{(2)}=3$ and ${\rank}\,G_2^{(2)}=4.$
Symmetries of lower ranks are trivial. 

An algorithm to compute generalized symmetries will be outlined in 
Section~\ref{algosymmetry} and applied to (\ref{todalattice}).
\begin{definition}
A recursion operator ${\mathcal R}$ connects symmetries 
\begin{equation}
\label{symmetrylink}
{\bf G}^{(j+s)} = {\mathcal R} {\bf G}^{(j)}, 
\end{equation}
where $ j=1,2,... ,$ and $s$ is the seed. 
In most cases the symmetries are consecutively linked $(s=1).$
For $N\/$-component systems, ${\mathcal R}$ is an $N \times N$ matrix operator.

The defining equation for ${\mathcal R}$ \cite{PO93,JW98} is
\begin{equation}
\label{definingrecursion}
{\rm D}_t {\mathcal R} + [{\mathcal R}, {\bf F}'({\bf u}_n) ] = 
\frac{\partial {\mathcal R}}{\partial t} + 
{\mathcal R}' [{\bf F}] + {\mathcal R} \circ {\bf F}'({\bf u}_n) - 
{\bf F}'({\bf u}_n) \circ {\mathcal R} = 0, 
\end{equation}
where $[\; , \; ]$ denotes the commutator and $\circ$ the composition of 
operators. 
The operator ${\bf F}'({\bf u}_n)$ was defined in 
(\ref{ddefrechetvectoroperator}).
${{\mathcal R}}' [{\bf F}]$ is the Fr\'echet derivative of 
${\mathcal R}$ in the direction of ${\bf F}.$ 
For the scalar case, the operator ${\mathcal R}$ is often of the form 
\begin{equation}
\label{recursionoperatorscalar}
{\mathcal R} = 
U(u_n) \, {\mathcal O}( ({\rm D} - {\rm I})^{-1}, 
{\rm D}^{-1}, {\rm I}, {\rm D}) \, V(u_n),
\end{equation}
and then
\begin{equation}
\label{frechetofrscalar}
{\mathcal R}' [F] = 
\sum_{k} \, ({\rm D}^k F) \, 
\frac{\partial U}{\partial u_{n+k}} \, {\mathcal O} \, V
+ \sum_{k} \, U {\mathcal O} \, ({\rm D}^k F) \,
\frac{\partial V}{\partial u_{n+k}}.
\end{equation}
For the vector case, the elements of the $N \times N$ operator matrix 
${\mathcal R}$ are often of the form 
${\mathcal R}_{ij} = 
U_{ij}({\bf u}_n) \, 
{\mathcal O}_{ij}( ({\rm D} -{\rm I})^{-1}, 
{\rm D}^{-1}, {\rm I}, {\rm D} ) \, V_{ij}({\bf u}_n). $
For the two-component case 
\begin{eqnarray}
\label{frechetofrvector}
{\mathcal R}'[{\bf F}]_{ij} &=& 
\sum_{k} \, ({\rm D}^k F_1) \, 
\frac{\partial U_{ij}}{\partial u_{n+k}} \, {\mathcal O}_{ij} \, V_{ij}
+ \sum_{k} \, ({\rm D}^k F_2) \, 
\frac{\partial U_{ij}}{\partial v_{n+k}} \, {\mathcal O}_{ij} \, V_{ij} 
\nonumber \\
&& +\! \sum_{k} \, U_{ij} {\mathcal O}_{ij} \, ({\rm D}^k F_1) \,
\frac{\partial V_{ij}}{\partial u_{n+k}}
+ \sum_{k} \, U_{ij} {\mathcal O}_{ij} \, ({\rm D}^k F_2) \,
\frac{\partial V_{ij}}{\partial v_{n+k}}.
\end{eqnarray}
\end{definition}
\begin{example}
The KvM lattice (\ref{kvmlattice}) has recursion operator
\begin{eqnarray}
\label{recursionkvm}
{\mathcal R} &=& u_n ({\rm I} + {\rm D})(u_n {\rm D} - {\rm D}^{-1} u_n )
({\rm D} - {\rm I})^{-1} \frac{1}{u_n} {\rm I} 
\nonumber \\
&=& u_n {\rm D}^{-1} + (u_n + u_{n+1}) {\rm I} + u_n {\rm D}
+ u_n ( u_{n+1} - u_{n-1} ) ({\rm D} - {\rm I})^{-1} \frac{1}{u_n} {\rm I}.
\end{eqnarray}
\end{example}
%
%
\begin{example}
The Toda lattice (\ref{todalattice}) has recursion operator
\begin{equation}
\label{recursiontoda}
{\mathcal R} = 
\left( \begin{array}{cc} 
-u_{n}{\rm I} & - {\rm D}^{-1}-{\rm I} 
+ (v_{n-1}-v_{n})({\rm D}-{\rm I})^{-1} \frac{1}{v_{n}} {\rm I} \\
-v_n {\rm I} - v_n {\rm D} 
& - u_{n+1}{\rm I} 
+ v_{n}(u_{n}-u_{n+1})({\rm D}-{\rm I})^{-1} \frac{1}{v_{n}} {\rm I}
\end{array} \right).
\end{equation}
\end{example}
In Section~\ref{algorecursionscalar} we will give an algorithm for the 
computation of scalar recursion operators like (\ref{recursionkvm}). 
In Section~\ref{algorecursionmatrix} we cover the matrix case and show how
(\ref{recursiontoda}) is computed. 
The algorithms complement those for recursion operators of PDEs 
presented in \cite{WHandUG99} and elsewhere in these proceedings 
\cite{DBandWHandJS04}.
%
%
We now introduce two powerful tools which will be used in the computation 
of densities and fluxes.
\section{The Discrete Variational Derivative (Euler Operator)}
\label{discreteeuler}

\begin{definition}
A function $E_n({\bf u}_n)$ is a total difference if there exists 
another function $J_n({\bf u}_n),$ such that 
$E_n = \Delta \, J_n = ({\rm D} - {\rm I}) \, J_n.$  
\end{definition}
\begin{theorem}
A necessary and sufficient condition for a function $E_n$ with
positive shifts up to level $p_0,$ to be a total difference is that
\begin{equation} 
\label{Econdition}
{\mathcal L}^{(0)}_{{\bf u}_n} (E_n) = 0,
\end{equation}
where ${\mathcal L}^{(0)}_{{\bf u}_n}$ is the discrete variational derivative
(Euler operator) \cite{VAandASandRY00}
defined by
\begin{equation}
\label{discreteeuleroperator}
{\mathcal L}^{(0)}_{{\bf u}_n} =
\frac{\partial}{\partial {\bf u}_{n}} ( \sum_{k=0}^{p_0} {\rm D}^{-k} )
= \frac{\partial}{\partial{{\bf u}_{n}}} 
( {\rm I} + {\rm D}^{-1} + {\rm D}^{-2} + \cdots + {\rm D}^{-p_0}).
\end{equation}
\end{theorem}
A proof is given in e.g.\ \cite{MHandWH03}.
\begin{remark}
To verify that an expression $E(u_{n-q}, \cdots, u_n, \cdots, u_{n+p})$ 
involving negative shifts is a total difference, one must first remove the 
negative shifts by replacing $E_n$ by ${\tilde E}_n = {\rm D}^q E_n.$
Applied to ${\tilde E}_n,$ (\ref{discreteeuleroperator}) terminates 
at $p_0 = p+q.$ 
\end{remark}
We now introduce a tool to invert the total difference operator 
$\Delta = {\rm D} - {\rm I}.$
\section{The Discrete Homotopy Operator}
\label{discretehomotopy}
Given is an expression $E_n$ (free of negative shifts). 
We assume that one has verified that 
$E_n \in {\rm Ker}\, {\mathcal L}^{(0)}_{{\bf u}_n},$ 
i.e.\ ${\mathcal L}^{(0)}_{{\bf u}_n} (E_n) = 0.$ 
Consequently, $E_n \in {\rm Im} \, \Delta.$ 
So, $\exists J_n$ such that $E_n = \Delta J_n.$
To compute $J_n = \Delta^{-1} (E_n)$ one must invert the operator 
$\Delta = {\rm D} - {\rm I}.$
Working with the formal inverse, 
\begin{equation}
\label{invdelta}
\Delta^{-1} = {\rm D}^{-1} + {\rm D}^{-2} + {\rm D}^{-3} + \ldots,
\end{equation}
is impractical, perhaps impossible. 
We therefore present the (discrete) homotopy operator 
which circumvents the above infinite series. 
In analogy to the continuous case \cite{PO93}, we first introduce the 
discrete higher Euler operators.
\begin{definition}
The discrete higher Euler operators are defined by
\begin{equation}
\label{discreteeulernew}
{\mathcal L}^{(i)}_{{\bf u}_n} = \frac{\partial}{\partial{{\bf u}_n}} 
(\sum_{k=i}^{p_i} {\binom ki} {\rm D}^{-k} )
\end{equation}
\end{definition}
The higher Euler operators all terminate at some maximal shifts $p_i.$
\begin{example}
For scalar component $u_n,$ the first higher Euler operators are:
\begin{eqnarray}
\label{highereuler0}
{\mathcal L}^{(0)}_{u_n} \!&\!=\!&\! 
\frac{\partial}{\partial {u_n}} 
( {\rm I} + {\rm D}^{-1} + {\rm D}^{-2} + 
{\rm D}^{-3} + \cdots + {\rm D}^{-p_0}), 
\\
\label{highereuler1}
{\mathcal L}^{(1)}_{u_n} \!&\!=\!&\! 
\frac{\partial}{\partial {u_n}} 
( {\rm D}^{-1} + 2 {\rm D}^{-2} + 3 {\rm D}^{-3} + 
4 {\rm D}^{-4} + \cdots + p_1 \,{\rm D}^{-p_1}),
\\
\label{highereuler2}
{\mathcal L}^{(2)}_{u_n} \!&\!=\!&\! 
\frac{\partial}{\partial{u_n}} 
( {\rm D}^{-2} + 3 {\rm D}^{-3} + 6 {\rm D}^{-4} + 
10 {\rm D}^{-5} + \cdots + \frac{1}{2} p_2 (p_{2}-1) \,{\rm D}^{-p_2}).
\end{eqnarray}
\end{example}
Note that (\ref{highereuler0}) coincides with (\ref{discreteeuleroperator}).
Similar formulae hold for ${\mathcal L}^{(i)}_{v_n}.$

Next, we introduce the homotopy operators.
For notational simplicity we show the formulae for the two component case 
${\bf u}_n = (u_n,v_n).$ 
\begin{definition}
The total homotopy operator is defined as
\begin{equation}
\label{totalhomotopyoperator}
{\mathcal H} = \int_{0}^{1} ( j_{1,n}({\bf u}_n)[\lambda {\bf u}_n] 
+ j_{2,n}({\bf u}_n)[\lambda {\bf u}_n] ) \frac{d \lambda}{\lambda},
\end{equation}
where the homotopy operators are given by
\begin{equation}
\label{homotopyoperator1and2}
j_{1,n}({\bf u}_n) = 
\sum_{i=0}^{{p_1}-1} 
({\rm D}-{\rm I})^i ( u_{n} {\mathcal L}^{(i+1)}_{u_{n}} ), \quad
j_{2,n}({\bf u}_n) =
\sum_{i=0}^{{p_2}-1} 
({\rm D}-{\rm I})^i ( v_{n} {\mathcal L}^{(i+1)}_{u_{n}} ).
\end{equation}
\end{definition}
%
%
\begin{theorem}
$J_n = \Delta^{-1} (E_n) $ can be computed as 
$J_n = {\mathcal H} (E_n).$
\end{theorem}
A similar theorem and proof for continuous homotopy operators is given 
in \cite{PO93}. 
By constructing a similar variational bicomplex the theorem still holds 
in the discrete case.
See \cite{PHandLM02,LMandPH02} and elsewhere in these proceedings 
\cite{EMandRQ04}.
\begin{remark}
Since the theory for the recursion operator can be formulated in the 
language of Lie algebra complexes, the results in \cite{JSandJW01a} 
are immediately applicable. 
This gives one explicit conditions from which the existence of infinite 
hierarchies of local symmetries and cosymmetries can be concluded.
Also, the problems with the definition of a recursion operator as treated 
in \cite{JSandJW01b} also play a role in the discrete case.
\end{remark}
\begin{remark}
Note that $p_1$ and $p_2$ in (\ref{homotopyoperator1and2}) are the 
highest shifts for $u_n$ and $v_n$ in $E_n.$ 
Furthermore, $j_{r,n} ({\bf u}_n)[\lambda {\bf u}_n]$ means 
that in $j_{r,n} ({\bf u}_n)$ 
one replaces ${\bf u}_n \rightarrow \lambda {\bf u}_n, $ and, of course,
$ {\bf u}_{n+1} \rightarrow \lambda {\bf u}_{n+1}, \;
{\bf u}_{n+2} \rightarrow \lambda {\bf u}_{n+2}, \; {\rm etc.} $
\end{remark}
\begin{remark}
In practice, one does not compute the definite integral 
(\ref{totalhomotopyoperator}) 
since the evaluation at boundary $\lambda=0$ may cause problems.
Instead, one computes the indefinite integral and evaluates the result at 
$\lambda=1.$
\end{remark}
%
%
\section{Algorithm to Compute Densities and Fluxes}
\label{algodensityflux}

As an example, we compute the density-flux pair (\ref{condenstoda3}) for 
(\ref{todalattice}).
Assuming that the weights (\ref{todaweights}) are computed and the rank
of the density is selected ($R=3$ here), the algorithm has three steps. 
\vskip 2.0pt
\noindent
{\bf Step 1: Construct the form of the density}
\vskip 2.0pt
\noindent
List all monomials in $u_n$ and $v_n$ of rank $3$ or less:
${\mathcal G} \!=\! \{ {u_n}^3, {u_n}^2, u_n v_n, {u_n}, v_n \}.$

Next, for each monomial in ${\mathcal G}$, introduce the correct number of
$t$-derivatives so that each term has weight $3.$ 
Thus, using (\ref{todalattice}), 
\begin{eqnarray*}
\label{todaweightadjust}
&&
\frac{{\rm d}^0}{{\rm d}t^0} ( {u_n}^3 ) = {u_n}^3 , 
\;\;\;\;\quad 
\frac{{\rm d}^0}{{\rm d}t^0} ( u_n v_n ) = u_n v_n , 
\nonumber \\
&&
\frac{{\rm d}}{{\rm d}t} ( {u_n}^2 ) = 
2 u_n {\dot{u}}_n = 2 u_n v_{n-1} - 2 u_n v_n ,  
\;\;\;\;\quad
\frac{{\rm d}}{{\rm d}t} ( v_n ) = 
{\dot{v}}_n =  u_n v_n -  u_{n+1} v_n, 
\nonumber \\
&&
\frac{{\rm d}^2}{{\rm d}t^2} ( u_n )= \frac{{\rm d}}{{\rm d}t} ( {\dot{u}}_n ) 
   = \frac{{\rm d}}{{\rm d}t} ( v_{n-1} - v_n ) 
   = u_{n-1} v_{n-1} - u_{n} v_{n-1} - u_n v_n + u_{n+1} v_n .
\end{eqnarray*}
Gather all terms in $\mathcal H=$
$\{ {u_n}^3, u_n v_{n-1} , u_n v_n , u_{n-1} v_{n-1} , u_{n+1} v_n \}. $
Identify members belonging to the same equivalence classes and 
replace them by their main representatives. 
For example, $ u_n v_{n-1} \equiv u_{n+1} v_n,$ so both are replaced 
by $u_n v_{n-1}.$ 
So, $\mathcal H$ is replaced by 
${\mathcal I} = \{ {u_n}^3 , u_n v_{n-1} , u_n v_n \}, $
which has the building blocks of the density. 
Linear combination of the monomials in ${\mathcal I}$ with constant 
coefficients $c_i$ gives 
\begin{equation}
\label{formrho3toda}
\rho_n = c_1 \, {u_n}^3 + c_2 \, u_n v_{n-1} + c_3 \, u_n v_n . 
\end{equation}
\vskip 0.1pt
\noindent
{\bf Step 2: Determine the coefficients}
\vskip 1.75pt
\noindent
Require that (\ref{ddeconslaw}) holds.
Compute ${\rm D}_t \rho_n.$ 
Use (\ref{todalattice}) to remove ${\dot{u}_n}$ and ${\dot{v}_n}$ 
and their shifts.
Thus, 
\begin{eqnarray}
\label{todalatticerhodot}
E_n \!&=&\! {\rm D}_t \rho_n =
( 3 c_1 - c_2 ) u_n^2 v_{n-1} + (c_3 - 3 c_1 ) u_n^2 v_n 
+ (c_3 - c_2) v_{n-1} v_n  
\nonumber \\
&& + c_2 u_{n-1} u_n v_{n-1} + c_2 v_{n-1}^2 -c_3 u_n u_{n+1} v_n - c_3 v_n^2.
\end{eqnarray}
Compute ${\tilde E}_n = {\rm D} E_n.$ 
First, apply (\ref{discreteeuleroperator}) for component $u_n$ to 
${\tilde E}_n:$ 
\begin{eqnarray}
\label{applyeulerforun}
{\mathcal L}^{(0)}_{u_n} ({\tilde E}_n) 
\!&\!=\!&\!
\frac{\partial}{\partial {u_{n}}}
( {\rm I} + {\rm D}^{-1} + {\rm D}^{-2}) ({\tilde E}_n) 
\nonumber \\
\!&\!=\!&\!
2 (3 c_1 - c_2 ) u_n v_{n-1} + 2 (c_3 - 3 c_1 ) u_n v_n 
\nonumber \\
&& + ( c_2 - c_3 ) u_{n-1} v_{n-1} + (c_2 - c_3) u_{n+1} v_{n}.
\end{eqnarray}
Second, apply (\ref{discreteeuleroperator}) for component $v_n$ to 
${\tilde E}_n:$ 
\begin{eqnarray}
\label{applyeulerforvn}
{\mathcal L}^{(0)}_{v_n} ({\tilde E}_n) 
\!&\!=\!&\!
\frac{\partial}{\partial {v_{n}}} ( {\rm I} + {\rm D}^{-1}) ({\tilde E}_n) 
\nonumber \\
\!&\!=\!&\!
(3 c_1 - c_2 ) u_{n+1}^2 + (c_3 - c_2) v_{n+1} 
+ ( c_2 - c_3 ) u_{n} u_{n+1} 
\nonumber \\
&& + 2 (c_2 - c_3) v_{n} + (c_3 - 3 c_1 ) u_n^2 + ( c_3 - c_2 ) v_{n-1}. 
\end{eqnarray}
Both (\ref{applyeulerforun}) and (\ref{applyeulerforvn}) must 
vanish identically. 
Solve the linear system
\begin{equation}
\label{todacsystem}
{\mathcal S} = \{ 3 c_1 - c_2 = 0, c_3 - 3 c_1 = 0, c_2 - c_3 = 0 \}. 
\end{equation}
The solution is $ 3 c_1 = c_2 = c_3.$ 
Substituting $c_1 = \frac{1}{3}, c_2 = c_3 = 1,$ into (\ref{formrho3toda})
\begin{equation}
\label{todacondens3final}
\rho_n = \frac{1}{3} {u_n}^3 + u_n ( v_{n-1} + v_n ).
\end{equation}
\vskip 0.01pt
\noindent
{\bf Step 3}: {\bf Compute the flux}
\vskip 2.0pt
\noindent
In view of (\ref{ddeconslaw}), we will compute 
$J_n = \Delta^{-1}(-E_n)$ with the homotopy operator introduced in 
Section~\ref{discretehomotopy}.
Alternative methods are described in \cite{UGandWH98,MHandWH03}.

Insert $c_1 = \frac{1}{3}, c_2 = c_3 =1 $ into (\ref{todalatticerhodot}) and 
compute
\begin{equation}
\label{todastartEtilde}
{\tilde E}_n = {\rm D} E_n = 
u_{n} u_{n+1} v_{n} + v_{n}^2 - u_{n+1} u_{n+2} v_{n+1} - v_{n+1}^2.
\end{equation}
\vskip 0.2pt
\noindent
We apply formulae (\ref{homotopyoperator1and2}) to $-{\tilde E}_n.$ 
The pieces are listed in Tables~\ref{todahomotopy1} and~\ref{todahomotopy2}.
%
\begin{center} 
\begin{table}[h] 
\begin{tabular}{|c|c|c|}\hline
$i$ 
 & ${\mathcal L}^{(i+1)}_{u_n} ( -{\tilde E}_n )$ 
 & $({\rm D}- {\rm I})^i ( u_n {\mathcal L}^{(i+1)}_{u_n} ( -{\tilde E}_n) )$ 
\\ \hline \hline
0 & $ u_{n-1} v_{n-1} \!+\! u_{n+1} v_{n} $ 
  & $ u_n u_{n-1} v_{n-1} \!+\! u_n u_{n+1} v_{n} $ 
\\
1 & $ u_{n-1} v_{n-1} $ 
  & $ u_{n+1} u_n v_{n} \!-\! u_{n} u_{n-1} v_{n-1} $ 
\\ \hline
\end{tabular}
\caption{Computation of ${j}_{1,n}({\bf u}_n)(-{\tilde E}_n) $ } 
\label{todahomotopy1}
\vspace{-0.75cm}
\end{table}
%
\begin{table}[h] 
\begin{tabular}{|c|c|c|}\hline
$i$ 
  & ${\mathcal L}^{(i+1)}_{v_n} ( -{\tilde E}_n )$ 
  & $({\rm D}- {\rm I})^i (v_n {\mathcal L}^{(i+1)}_{v_n} ( -{\tilde E}_n) )$ 
\\ \hline \hline
0 & $ u_{n} u_{n+1} \!+\! 2 v_{n} $ 
  & $ v_n u_{n} u_{n+1} \!+\! 2 v_{n}^2 $ 
\\ \hline
\end{tabular}
\caption{Computation of ${j}_{2,n}({\bf u}_n)(-{\tilde E}_n) $ } 
\label{todahomotopy2}
\vspace{-0.75cm}
\end{table}
\end{center}
Adding the terms in the right columns in Table~\ref{todahomotopy1}
and Table~\ref{todahomotopy2}, 
\begin{equation}
\label{todaj1andj2}
{j}_{1,n}({\bf u}_n)(-{\tilde E}_n) = 2 u_n u_{n+1} v_n, \quad
{j}_{2,n}({\bf u}_n)(-{\tilde E}_n) = u_n u_{n+1} v_n + 2 v_n^2.
\end{equation}
Thus, the homotopy operator (\ref{totalhomotopyoperator}) gives
\begin{eqnarray}
\label{todatotalhomotopy}
\tilde{J}_n &=& 
  \int_0^1 ( {j}_{1,n}({\bf u}_n)(-{\tilde E}_n)[\lambda {\bf u}_n] + 
  {j}_{2,n}({\bf u}_n)(-{\tilde E}_n)[\lambda {\bf u}_n]) \; 
\frac{d\lambda}{\lambda} 
\nonumber \\ 
&=&  \int_0^1 ( 3 \lambda^2 u_n u_{n+1} v_n + 2 \lambda v_n^2 ) \; d\lambda  
\nonumber \\
&=& u_n u_{n+1} v_n + v_n^2.
\end{eqnarray}
After a backward shift, $J_n = {\rm D}^{-1} {\tilde J}_n,$ we obtain 
the final result:
\begin{equation} 
\label{finalrhoandj}
\rho_n = \frac{1}{3} \, u_n^3 + u_n ( v_{n-1} + v_n ),
\quad 
J_n = u_{n-1} u_n v_{n-1} + v_{n-1}^2.
\end{equation}
\section{Algorithm to Compute Generalized Symmetries}
\label{algosymmetry}

As an example, we compute the symmetry (\ref{todasymm2}) of rank $(3,4)$ 
for (\ref{todalattice}).
The two steps of the algorithm \cite{UGandWH99} are similar to those in 
Section~\ref{algodensityflux}.
\vskip 2pt
\noindent
{\bf Step 1: Construct the form of the symmetry}
\vskip 2pt
\noindent
Start by listing all monomials in $u_n$ and $v_n$ of ranks $3$ and 4, or less: 
\begin{equation}
\label{listmonomialsymtoda}
{\mathcal K}_1 = \{ u_n^3, u_n^2, u_n v_n, u_n, v_n \}, \quad
{\mathcal K}_2 = 
\{ u_n^4, u_n^3, u_n^2 v_n, u_n^2, u_n v_n, u_n, v_n^2, v_n \}.
\nonumber 
\end{equation}
%
As in Step 1 in Section~\ref{algodensityflux}, 
for each monomial in ${\mathcal K}_1$  and ${\mathcal K}_2$, introduce 
the necessary $t$-derivatives so that each term has rank $3$ and $4$, 
respectively. 
At the same time, use (\ref{todalattice}) to remove all $t-$derivatives.
Doing so, based on ${\mathcal K}_1$ we obtain
\begin{equation}
\label{setr1toda}
{\mathcal L}_1 = 
\{ u_n^3, u_{n-1} v_{n-1} , u_n v_{n-1} , u_n v_n , u_{n+1} v_n \} .
\end{equation}
\vskip 2pt
\noindent
Similarly, based on ${\mathcal K}_2,$ we get
\begin{eqnarray}
\label{setr2toda}
{\mathcal L}_2 & = &
\{ u_n^4, u_{n-1}^2 v_{n-1}, u_{n-1} u_n v_{n-1}, u_n^2 v_{n-1},
v_{n-2} v_{n-1}, v_{n-1}^2, u_n^2 v_n, 
\nonumber \\
& &  u_n u_{n+1} v_n, u_{n+1}^2 v_n,
v_{n-1} v_n, v_n^2, v_n v_{n+1} \} .
\end{eqnarray}
In contrast to the strategy for densities, we do not introduce the main 
representatives, but linearly combine the monomials in 
${\mathcal L}_1$ and ${\mathcal L}_2$ to get the form of the symmetry:
\begin{eqnarray}
\label{formsym3toda}
G_1 &=& c_1 \, u_n^3 + c_2 \, u_{n-1} v_{n-1} + c_3 \, u_n v_{n-1} +
c_4 \, u_n v_n + c_5 \, u_{n+1} v_n, \nonumber \\
G_2 &=& c_6 \, u_n^4 + c_7 \, u_{n-1}^2 v_{n-1} + 
c_8 \, u_{n-1} u_n v_{n-1} + c_9 \, u_n^2 v_{n-1} +
c_{10} \, v_{n-2} v_{n-1} \nonumber \\ 
&& + c_{11} \, v_{n-1}^2 + c_{12} \, u_n^2 v_n +
c_{13} \, u_n u_{n+1} v_n + c_{14} \, u_{n+1}^2 v_n +
c_{15} \, v_{n-1} v_n \nonumber \\
&& + c_{16} \, v_n^2 + c_{17} \, v_n v_{n+1}, 
\end{eqnarray} 
with constant coefficients $c_i.$
\vskip 3pt
\noindent
{\bf Step 2: Determine the coefficients}
\vskip 2pt
To compute the coefficients $c_i$ we require that (\ref{ddesymmetry}) holds 
on solutions of (\ref{dde}).
Compute ${\rm D}_t G_1$ and ${\rm D}_t G_2$ and use (\ref{todalattice}) to 
remove ${\dot{u}}_{n}, {\dot{v}}_n,$ and their shifts. 
This gives the left hand sides of (\ref{ddesymmetry}). 

Use (\ref{ddefrechetcomponent}) to get the right hand sides of 
(\ref{ddesymmetry}): 
\begin{eqnarray}
\label{todafrechetcomponent1and2}
{F_1}'({\bf u}_n)[{\bf G}] &=& {\rm D}^{-1} G_2 - {\rm I} G_2, 
\nonumber \\
{F_2}'({\bf u}_n)[{\bf G}] &=& 
v_n {\rm I} G_1 - v_n {\rm D} G_1 + (u_n - u_{n+1}) {\rm I} G_2.
\end{eqnarray}
Substitute (\ref{formsym3toda}) into (\ref{todafrechetcomponent1and2}) and
equate the corresponding left and right hand sides.
Since all monomials in $u_n, v_n$ and their shifts are independent, 
one obtains the linear system that determines the coefficients $c_i.$
The solution is
\begin{eqnarray}
\label{sym3todaresult}
c_1 = c_6 = c_7 = c_8 = c_9 = c_{10} = c_{11} = c_{13} = c_{16} = 0, 
\nonumber \\
c_2 = c_3 = -c_4 = -c_5 = c_{12} = -c_{14} = c_{15} = -c_{17} .
\end{eqnarray}
With the choice $c_{17}=1,$ the symmetry (\ref{formsym3toda}) finally becomes
\begin{eqnarray}
\label{todasym34}
G_1 &=& 
v_n (u_n  + u_{n+1}) - v_{n-1} (u_{n-1} - u_{n}) , 
\nonumber \\
G_2 &=& 
v_n (u_{n+1}^2 - u_n^2 + v_{n+1} - v_{n-1}).  
\end{eqnarray}
%
\section{Algorithm to Compute Scalar Recursion Operators}
\label{algorecursionscalar}
In this section we show how to compute the recursion operator
(\ref{recursionkvm}) of (\ref{todalattice}).

Again, we will use the concept of rank invariance to construct a candidate 
recursion operator. 
The defining equation (\ref{definingrecursion}) is then used to determine 
the coefficients. 

We observe that (\ref{recursionkvm}) in expanded form naturally splits 
into two pieces: 
\begin{equation}
\label{recursionsplit}
{\mathcal R} = {\mathcal R}_0 + {\mathcal R}_1, 
\end{equation}
where ${\mathcal R}_{0}$ contains only terms with shift operators
${\rm D}^{-1}, {\rm I},$ and ${\rm D}$, and ${\mathcal R}_{1}$ 
has terms involving $({\rm D} - {\rm I})^{-1}.$ 
\vskip 3pt
\noindent
{\bf Step 1: Determine the rank of the recursion operator}
\vskip 2pt
\noindent
In view of (\ref{symmetrylink}) and assuming that the symmetries are linked 
consecutively $(s=1)$, the recursion operator ${\mathcal R}$ has rank
\begin{equation}
\label{rankrecursionkvm}
R = {\rm rank} \; {\mathcal R} = 
{\rm rank}\; G^{(2)} - {\rm rank}\; G^{(1)} = 3 - 2 = 1, 
\end{equation}
where we used (\ref{kvmsymm1}) and (\ref{kvmsymm2}).
If the assumption turns out to be correct, then the recursion operator 
has rank 1.
If the assumption were falls because symmetries are not linked 
consecutively, then $R$ must be adjusted and the subsequent 
steps must be repeated.
See \cite{DBandWHandJS04} for examples of PDEs for which that happens.
\vskip 2pt
\noindent
{\bf Step 2: Determine the form of the recursion operator}
\vskip 2pt
\noindent
We split this into two sub-steps.
\vskip 1pt
\noindent
{\bf (i)} $\;$ {\bf Determine the form of ${\mathcal R}_{0}$}
\vskip 1pt
\noindent
The candidate ${\mathcal R}_{0}$ is a sum of terms involving 
${\rm D}^{-1}, {\rm I},$ and ${\rm D}.$ 
The coefficients of these terms are linear combinations of 
$u_{n-1}, u_{n},$ and $u_{n+1},$ so that all terms have the correct rank.
\begin{eqnarray}
\label{r0candidatekvm}
{\mathcal R}_{0} 
&=& (c_{1}u_{n-1} + c_{2}u_{n} + c_{3}u_{n+1})\, {\rm D}^{-1}
+ (c_{4}u_{n-1} + c_{5}u_{n} + c_{6}u_{n+1})\, {\rm I}
\nonumber \\ 
& & + (c_{7}u_{n-1} + c_{8}u_{n} + c_{9}u_{n+1})\, {\rm D},
\end{eqnarray}
where the $c_{i}$ are constant coefficients.
%

A few remarks are in place.
First, in ${\mathcal R}_0$ we moved the operators 
${\rm D}^{-1}, {\rm I},$ and ${\rm D}$ all the way to the right.
Second, the maximum up-shift and down-shift operator that should be
included can be determined by comparing two consecutive symmetries. 
Indeed, if the maximum up-shift in the first symmetry is $u_{n+p},$ and 
the maximum up-shift in the next symmetry is $u_{n+p+r},$ then 
${\mathcal R}_{0}$ must have ${\rm D}, {\rm D}^2, \cdots, {\rm D}^{r}.$
The same line of reasoning determines the minimum down-shift operator to 
be included.
In our example, there is no need to include terms in 
${\rm D}^{-2}, {\rm D}^2,$ etc.
Third, the coefficients of the operators can be restricted to linear
combinations of the terms appearing in $F.$ 
Hence, no terms in $u_{n \pm 2}, u_{n \pm 3}$ and so on occur in 
(\ref{r0candidatekvm}).
\vskip 2pt
\noindent
{\bf (ii)} $\;$ {\bf Determine the form of ${\mathcal R}_1$}
\vskip 1pt
\noindent
As in the continuous case \cite{WHandUG99}, 
${\mathcal R}_{1}$ is a linear combination (with constant coefficients 
${\tilde c}_{jk})$ of sums of all suitable products of symmetries 
and covariants (Fr\'echet derivatives of densities) sandwiching 
$({\rm D}-{\rm I})^{-1},$ i.e.\  
\begin{equation}
\label{formr1operator}
{\mathcal R}_{1} = 
\sum_j \sum_k {\tilde c}_{jk} G^{(j)} ({\rm D}-{\rm I})^{-1} \rho_n^{(k)'}.
\end{equation}
For (\ref{kvmlattice}), $G^{(1)}$ in (\ref{kvmsymm1}) and 
$\rho_n^{(0)} = \ln(u_n)$ in (\ref{kvmconslaw1}) are the only suitable pair.
Indeed, using (\ref{ddefrechetscalaroperator}) we have 
$\rho_n^{(0)'} = \ln(u_n)' = \frac{1}{u_n} {\rm I},$ which has rank -1. 
Combined with $G^{(1)}$ of rank 2, we have a term of rank 1.
Other combinations of symmetries and covariants would exceed rank 1.
Therefore, 
\begin{equation}
\label{r1candidatekvm}
{\mathcal R}_{1} 
= {\tilde c}_{10}u_{n}(u_{n+1}-u_{n-1})({\rm D}-{\rm I})^{-1}
(\frac{1}{u_{n}}) {\rm I}
\end{equation}
where ${\tilde c}_{10}$ a constant coefficient.
Using (\ref{recursionsplit}) and renaming ${\tilde c}_{10}$ to $c_{10},$
\begin{eqnarray}
\label{rcandidatekvm}
{\mathcal R} &=& 
(c_{1}u_{n-1} + c_{2}u_{n} + c_{3}u_{n+1}){\rm D}^{-1}
+ (c_{4}u_{n-1} + c_{5}u_{n} + c_{6}u_{n+1}){\rm I}
\nonumber \\ 
&& +
(c_{7}u_{n-1} + c_{8}u_{n} + c_{9}u_{n+1}){\rm D} 
+ c_{10}u_{n}(u_{n+1}-u_{n-1})({\rm D}-{\rm I})^{-1}
(\frac{1}{u_{n}}){\rm I}
\end{eqnarray}
is the candidate recursion operator for (\ref{kvmlattice}).
\vskip 3pt
\noindent
{\bf Step 3: Determine the coefficients}
\vskip 2pt
\noindent
Starting from (\ref{rcandidatekvm}), we use (\ref{frechetofrscalar})
with $F = u_n (u_{n+1} - u_{n-1})$ to compute ${\mathcal R}'[F].$ 
The partial results will not be shown due to length.

Using (\ref{ddefrechetscalar}), we compute
\begin{equation}
\label{frechetoffkvm}
F' = -u_{n} {\rm D}^{-1} + (u_{n+1}-u_{n-1}) {\rm I} +u_{n} {\rm D}.
\end{equation}
Then we compute ${\mathcal R} \circ F'$ and $F' \circ {\mathcal R}.$ 
After substituting the pieces into (\ref{definingrecursion}) we simplify 
the resulting expression using rules such as
\begin{eqnarray}
\label{scalardifferencerules}
({\rm D}-{\rm I})^{-1} {\rm D} 
&\!=\!& 
{\rm D} ({\rm D}-{\rm I})^{-1} 
= {\rm I} + ({\rm D}-{\rm I})^{-1},
\nonumber \\
({\rm D}-{\rm I})^{-1} {\rm D}^{-1} 
&\!=\!& 
{\rm D}^{-1} ({\rm D}-{\rm I})^{-1} 
= -{\rm D}^{-1} + ({\rm D}-{\rm I})^{-1}.
\end{eqnarray}
We further simplify by recursively using formulas like
\begin{eqnarray}
\label{scalarrecursiveformula}
\;\;\;\;\;\quad 
{\rm D} U(u_n)({\rm D}\!-\!{\rm I})^{-1} V(u_n) {\rm I} 
\!&\!\!\!=\!\!\!&\!U(u_{n+1}) V(u_n) {\rm I} \!+\! U(u_{n+1}) 
({\rm D}\!-\!{\rm I})^{-1} V(u_n) {\rm I},
\nonumber \\
\;\;\;\;\;({\rm D}\!-\!{\rm I})^{-1} U(u_n) V(u_n) {\rm D} 
\!&\!\!\!=\!\!\!&\!U(u_{n-1}) V(u_{n-1}) {\rm I} \!+\!
({\rm D}\!-\!{\rm I})^{-1} U(u_{n-1}) V(u_{n-1}) {\rm I}.
\end{eqnarray}
Finally, we equate like terms to obtain a linear system for the $c_i.$ 
Substituting the solution
\begin{equation}
\label{solutioncirecursionkvm}
c_{1} = c_{3} = c_{4} = c_{7} = c_{9} = 0, \quad c_{2} = c_{5} = c_{6} 
= c_{8} = c_{10} = 1
\end{equation}
into (\ref{rcandidatekvm}) we obtain the final result 
\begin{equation}
\label{kvmrecursionoperatorrepeat}
{\mathcal R} 
= u_n {\rm D}^{-1} + (u_n + u_{n+1}) {\rm I} + u_n {\rm D}
+ u_n ( u_{n+1} - u_{n-1} ) ({\rm D} - {\rm I})^{-1} \frac{1}{u_n} {\rm I}.
\end{equation}
A straightforward computation confirms that 
${\mathcal R} G^{(1)} = G^{(2)}$ with $G^{(1)}$ in (\ref{kvmsymm1}) and 
$G^{(2)}$ in (\ref{kvmsymm2}).
\section{Algorithm to Compute Matrix Recursion Operators}
\label{algorecursionmatrix}
We construct the recursion operator (\ref{recursiontoda}) for 
(\ref{todalattice}). 
Now all the terms in (\ref{definingrecursion}) are $2 \times 2$ 
matrix operators.
\vskip 3pt
\noindent
{\bf Step 1: Determine the rank of the recursion operator}
\vskip 2pt
\noindent
The difference in rank of symmetries is again used to compute the rank of 
the elements of the recursion operator.
Using (\ref{todaweights}), (\ref{todasymm1}) and (\ref{todasymm2}), 
\begin{equation}
\label{rankrtoda}
%
{\rm rank} \, {\bf G}^{(1)}
= \left( \begin{array}{c}
2 \\ 
3
\end{array} \right), 
\quad
{\rm rank}\, {\bf G}^{(2)}
= \left( \begin{array}{c}
3 \\ 
4
\end{array} \right).
\end{equation}
Assuming that ${\mathcal R} \, {\bf G}^{(1)} = {\bf G}^{(2)},$ 
we use the formula
\begin{equation}
\label{recursionrankrules}
{\rm rank}\, {\mathcal R}_{ij}
= {\rm rank}\, G^{(k+1)}_{i} - {\rm rank}\, G^{(k)}_{j}, 
\end{equation}
to compute a rank matrix associated to the operator
\begin{equation}
\label{todarankmatrix}
{\rm rank} \, {\mathcal R} 
= \left( \begin{array}{cc}
1 & 
0
\\
2 & 
1
\end{array} \right).
\end{equation}
\vskip 0.5pt
\noindent
{\bf Step 2: Determine the form of the recursion operator}
\vskip 3pt
\noindent
As in the scalar case, we build a candidate ${\mathcal R}_{0}:$ 
\begin{eqnarray}
\label{r0candidatetoda}
{\mathcal R}_{0} 
&=& \left( \begin{array}{cc}
({\mathcal R}_{0})_{11} & ({\mathcal R}_{0})_{12} \\
({\mathcal R}_{0})_{21} & ({\mathcal R}_{0})_{22}
\end{array} \right), 
\end{eqnarray}
with
\begin{eqnarray}
({\mathcal R}_{0})_{11} 
&=& (c_{1}u_{n} + c_{2}u_{n+1})\, {\rm I},
\nonumber \\
({\mathcal R}_{0})_{12} 
&=& c_{3}{\rm D}^{-1} + c_{4}{\rm I},
\nonumber \\
({\mathcal R}_{0})_{21} 
&=& (c_{5}u_{n}^{2} + c_{6}u_{n}u_{n+1} + c_{7}u_{n+1}^{2}
+ c_{8}v_{n-1} + c_{9}v_{n})\, {\rm I}
\nonumber \\ 
&& + (c_{10}u_{n}^{2} + c_{11}u_{n}u_{n+1} + c_{12}u_{n+1}^{2}
+ c_{13}v_{n-1} + c_{14} v_{n})\, {\rm D},
\nonumber \\
({\mathcal R}_{0})_{22} &=& (c_{15}u_{n} + c_{16}u_{n+1})\, {\rm I}.
\nonumber
\end{eqnarray}
Analogous to the scalar case, the elements of matrix ${\mathcal R}_{1}$ 
are linear combinations with constant coefficients of all suitable products 
of symmetries and covariants sandwiching $({\rm D}-{\rm I})^{-1}.$
Hence,
\begin{equation}
\label{operatorr1}
\sum_j \sum_k {\tilde c}_{jk} {\bf G}^{(j)} ({\rm D}-{\rm I})^{-1} 
\otimes \rho_n^{(k)'}, 
\end{equation}
where $\otimes$ denotes the matrix outer product, defined as
\[
\label{outerproduct}
\left( \begin{array}{c}
\!G_1^{(j)}\!\\ 
\!G_2^{(j)}\!\end{array} \right) 
({\rm D}\!-\!{\rm I})^{-1} \otimes
\left( \rho_{n,1}^{(k)'}\;\;\rho_{n,2}^{(k)'} \right)
= \left( \begin{array}{cc}
G_1^{(j)} ({\rm D}\!-\!{\rm I})^{-1}\rho_{n,1}^{(k)'} & 
G_1^{(j)} ({\rm D}\!-\!{\rm I})^{-1}\rho_{n,2}^{(k)'}
\\
G_2^{(j)} ({\rm D}\!-\!{\rm I})^{-1} \rho_{n,1}^{(k)'} & 
G_2^{(j)} ({\rm D}\!-\!{\rm I})^{-1} \rho_{n,2}^{(k)'}
\end{array} \right).
\]
%
Only the pair
$({\bf G}^{(1)},\rho_n^{(0)'})$ can be used, otherwise the ranks in 
(\ref{todarankmatrix}) would be exceeded. 
Using (\ref{ddefrechetcomponent}) and (\ref{condenstoda0}) we compute 
\begin{equation}
\rho_n^{(0)'} = 
\left( \begin{array}{cc} {\rm 0} & \frac{1}{v_n} {\rm I} \end{array} \right), 
\end{equation}
Therefore, using (\ref{operatorr1}) and renaming ${\tilde c}_{10}$ to $c_{17},$
\begin{equation}
\label{r1candidatetoda}
{\mathcal R}_{1} 
= \left( \begin{array}{cc}
{\rm 0} \, &
c_{17}(v_{n-1}-v_{n})({\rm D}\!-\!{\rm I})^{-1}\, \frac{1}{v_{n}}\, {\rm I}
\\ 
{\rm 0} \, & 
c_{17}v_{n}(u_{n}-u_{n+1})({\rm D}\!-\!{\rm I})^{-1}\, 
\frac{1}{v_{n}}\, {\rm I}
\end{array} \right).
\end{equation}
%
%
Adding (\ref{r0candidatetoda}) and (\ref{r1candidatetoda}) we obtain 
\begin{eqnarray}
\label{rcandidatetoda}
{\mathcal R} 
&=& \left( \begin{array}{cc}
{\mathcal R}_{11} & {\mathcal R}_{12} \\
{\mathcal R}_{21} & {\mathcal R}_{22}
\end{array} \right), 
\end{eqnarray}
with
\begin{eqnarray}
{\mathcal R}_{11} 
&=& (c_{1}u_{n} + c_{2}u_{n+1})\, {\rm I} 
\nonumber \\
{\mathcal R}_{12} 
&=& c_{3}{\rm D}^{-1} + c_{4}{\rm I}
+ c_{17}(v_{n-1}-v_{n})({\rm D}-{\rm I})^{-1}\, \frac{1}{v_{n}} \,{\rm I},
\nonumber \\
{\mathcal R}_{21} 
&=& (c_{5}u_{n}^{2} + c_{6}u_{n}u_{n+1} + c_{7}u_{n+1}^{2}
    + c_{8}v_{n-1} + c_{9}v_{n})\, {\rm I}
\nonumber \\ 
& & + (c_{10}u_{n}^{2} + c_{11}u_{n}u_{n+1} + c_{12}u_{n+1}^{2}
    + c_{13}v_{n-1} + c_{14}v_{n})\, {\rm D},
\nonumber \\
{\mathcal R}_{22} 
&=& (c_{15}u_{n} + c_{16}u_{n+1}){\rm I}
    + c_{17}v_{n}(u_{n}-u_{n+1})({\rm D}-{\rm I})^{-1}\, 
    \frac{1}{v_{n}} \,{\rm I}.
\nonumber
\end{eqnarray}
\vfill
\newpage
\noindent
{\bf Step 3: Determine the coefficients}
\vskip 2pt
\noindent
All the terms in (\ref{definingrecursion}) need to be computed.
The strategy is similar to the scalar case, yet the computations are 
much more cumbersome.
Omitting the details, the result is:
$ c_{2} = c_{5} = c_{6} = c_{7} = c_{8} = c_{10} = c_{11} 
= c_{12} = c_{13} = c_{15} = 0, \quad
c_{1} = c_{3} = c_{4} = c_{9} = c_{14} = c_{16} = -1,$ and $c_{17} = 1.$
Substitution the constants into (\ref{rcandidatetoda}) gives 
\begin{equation}
\label{recursionrtoda}
{\mathcal R} 
= \left( \begin{array}{cc} 
-u_{n}{\rm I} 
& 
- {\rm D}^{-1}-{\rm I} 
+ (v_{n-1}-v_{n})({\rm D}-{\rm I})^{-1}\, \frac{1}{v_{n}}\, {\rm I} \\ 
- v_n{\rm I}-v_n{\rm D}
& 
- u_{n+1}{\rm I} 
+ v_{n}(u_{n}-u_{n+1})({\rm D}-{\rm I})^{-1}\, \frac{1}{v_{n}} \,{\rm I}
\end{array} \right).
\end{equation}
It is straightforward to verify that ${\mathcal R} G^{(1)} = G^{(2)}$ 
with $G^{(1)}$ in (\ref{todasymm1}) and $G^{(2)}$ in  (\ref{todasymm2}). 
\section{More Examples}
\label{examples}
\begin{example}
The modified Volterra lattice \cite{VAandASandRY00,MHandWH03},
\begin{equation}
\label{modifiedvolterralattice}
{\dot {u}}_n = u_n^2 (u_{n+1} - u_{n-1}),
\end{equation}
has two non-polynomial densities $\rho_n^{(0)} = \frac{1}{u_n}$  
and $\rho_n^{(1)} = \ln(u_n),$ and infinitely many polynomial densities.
The first two symmetries, 
\begin{eqnarray}
\label{modifiedvolterrasymm1}
G^{(1)} &=& u_n^2 (u_{n+1} - u_{n-1}), \\
\label{modifiedvolterrasymm2}
G^{(2)} &=& u_n^2 u_{n+1}^2 (u_n + u_{n+2})
- u_{n-1}^2 u_{n}^2 (u_{n-2} + u_n),
\end{eqnarray}
are linked by the recursion operator 
\begin{equation}
\label{recursionmodifiedvolterra}
{\mathcal R} =
u_n^2 {\rm D}^{-1} + 2 u_n u_{n+1} {\rm I} + u_n^2 {\rm D} +
2 u_n^2 ( u_{n+1} - u_{n-1} ) ({\rm D}-{\rm I})^{-1} \frac{1}{u_n} {\rm I}.
\end{equation}
\end{example}
\begin{example}
The AL lattice (\ref{ablowitzladiklattice}) has infinitely 
many densities \cite{UGandWH98} and symmetries \cite{UGandWH99}.
The recursion operator is of the form (\ref{rcandidatetoda}) with
\begin{eqnarray}
\label{recursionablowitzladikelements}
{\mathcal R}_{11} 
&=& 
P_n {\rm D}^{-1} - u_{n} \Delta^{-1} v_{n+1} {\rm I}
- u_{n-1} P_n \Delta^{-1} \frac{v_n}{P_n} {\rm I},
\nonumber \\
{\mathcal R}_{12} 
&=& 
- u_n u_{n-1} {\rm I} - u_n \Delta^{-1} u_{n-1} {\rm I}
- u_{n-1} P_n \Delta^{-1} \frac{u_n}{P_n} {\rm I},
\nonumber \\
{\mathcal R}_{21} 
&=& 
v_n v_{n+1} {\rm I} + v_n \Delta^{-1} v_{n+1} {\rm I}
+ v_{n+1} P_n \Delta^{-1} \frac{v_n}{P_n} {\rm I},
\nonumber \\
{\mathcal R}_{22} 
&=& 
( u_n v_{n+1} + u_{n-1} v_n ) {\rm I} + P_n {\rm D} 
+ v_n \Delta^{-1} v_{n-1} {\rm I}
+ v_{n+1} P_n \Delta^{-1} \frac{u_n}{P_n} {\rm I}, 
\end{eqnarray}
where $P_n = 1 + u_n v_n$ and $\Delta = {\rm D}-{\rm I}.$
\end{example}
This recursion operator has an inverse, which is quite exceptional. 
The investigation of the recursion operator structure of the 
AL lattice is work in progress.
\vskip 2pt
\noindent
%
%
{\bf Acknowledgements} 
\vskip 2pt
\noindent
M.\ Hickman and B.\ Deconinck are gratefully acknowledged for valuable
discussions. D.\ Baldwin is thanked for proof reading the manuscript.
%
%
%
%
%
%
\bibliographystyle{amsalpha}

\end{document}